%% file: main.tex
\documentclass[sigconf]{acmart}

\copyrightyear{2022}
\acmYear{2022}
\setcopyright{acmcopyright}
\acmConference[KDD '22] {Proceedings of the 28th ACM SIGKDD Conference on Knowledge Discovery and Data Mining}{August 14--18, 2022}{Washington, DC, USA.}
\acmBooktitle{Proceedings of the 28th ACM SIGKDD Conference on Knowledge Discovery and Data Mining (KDD '22), August 14--18, 2022, Washington, DC, USA}
\acmPrice{15.00}
\acmISBN{978-1-4503-9385-0/22/08}
\acmDOI{10.1145/3534678.3539269}

\usepackage{booktabs} 
\usepackage{enumitem}
\usepackage{bm}
\usepackage{verbatim}

\usepackage{algorithm}
\usepackage{algorithmic}
\usepackage{booktabs} 
\usepackage{amsfonts}
\usepackage{amsmath}
\usepackage{setspace}
\usepackage{multirow}
\usepackage{subfigure}
\usepackage{enumitem}
\usepackage{bm}
\usepackage{verbatim}
\usepackage{makecell}
\usepackage{graphicx}
\usepackage{flushend}

\newcommand{\hide}[1]{}
\newcommand{\model}{{{\tt CLOVER}}}
\newcommand{\std}[1]{\tiny{$\pm$#1}}
\begin{document}



\title{Comprehensive Fair Meta-learned Recommender System}



\author{Tianxin Wei}
\affiliation{\institution{University of Illinois at Urbana Champaign}
\country{USA}}
\email{twei10@illinois.edu}

\author{Jingrui He}
\affiliation{\institution{University of Illinois at Urbana Champaign}
\country{USA}}
\email{jingrui@illinois.edu}

\begin{spacing}{0.96}
\input{0-abs}
\input{0-ccs_keywords}
\maketitle
\input{1-introduction}
\input{3-methodology}
\input{4-experiments}

\input{2-relatedwork}
\input{5-conclusion}
\begin{acks}
This work is supported by National Science Foundation under Award No. IIS-1947203, IIS-2117902, IIS-2137468, and Agriculture and Food Research Initiative (AFRI) grant no. 2020-67021-32799/project accession no.1024178 from the USDA National Institute of Food and Agriculture. The views and conclusions are those of the authors and should not be interpreted as representing the official policies of the funding agencies or the government.
\end{acks}
\bibliographystyle{ACM-Reference-Format}
\bibliography{ref}
\flushend
\input{6-appendix}

\end{spacing}
\end{document}

%% file: 0-abs.tex
\begin{abstract}

In recommender systems, one common challenge is the cold-start problem, where interactions are very limited for fresh users in the systems. To address this challenge, recently, many works introduce the meta-optimization idea into the recommendation scenarios, i.e. learning to learn the user preference by only a few past interaction items. The core idea is to learn global shared meta-initialization parameters for all users and rapidly adapt them into local parameters for each user respectively. They aim at deriving general knowledge across preference learning of various users, so as to rapidly adapt to the future new user with the learned prior and a small amount of training data. However, previous works have shown that recommender systems are generally vulnerable to bias and unfairness. Despite the success of meta-learning at improving the recommendation performance with cold-start, the fairness issues are largely overlooked.

In this paper, we propose a comprehensive fair meta-learning framework, named \textit{\model}, for ensuring the fairness of meta-learned recommendation models. We systematically study three kinds of fairness - individual fairness, counterfactual fairness, and group fairness in the recommender systems, and propose to satisfy all three kinds via a multi-task adversarial learning scheme. Our framework offers a generic training paradigm that is applicable to different meta-learned recommender systems. We demonstrate the effectiveness of \model~on the representative meta-learned user preference estimator on three real-world data sets. Empirical results show that \model~achieves comprehensive fairness without deteriorating the overall cold-start recommendation performance.


\end{abstract}

%% file: 0-ccs_keywords.tex
\ccsdesc[500]{Information systems~Recommender systems}

\keywords{Fairness, Recommender Systems, Meta-Learning}

%% file: 1-introduction.tex

\section{Introduction}
Personalized recommender systems have been widely used for mining user preference in various web services, such as e-commerce \cite{ying2018graph}, search engines \cite{shen2005implicit}, and social media \cite{fan2019graph}, which largely relieve the information overload issues. In practice, the common challenge that most recommender systems face is the cold-start problem. When data is scarce or fresh users appear frequently, the recommender systems must adapt rapidly. Over the past years, meta-learning methods \cite{lu2020meta,wei2020fast,dong2020mamo,lee2019melu} have been widely used to create recommender systems that learn quickly from the limited data with computationally affordable fine-tuning. The core idea of meta-learning is learning-to-learn, i.e., learning to solve the training tasks well with the generalization ability on future tasks. Meta-learned recommender systems are trained with an interleaving training procedure, comprised of inner loop update that fine-tunes on each user and outer loop update that produces the initialization of all users. However, recommender systems are generally vulnerable to bias and unfairness \cite{li2021towards,wu2021learning}. Although meta-learning could better help fresh users to find potentially interesting items, the understanding and mitigation of fairness in this context is largely under-explored.

\begin{figure}[t]
  \centering
  \includegraphics[width=0.9\linewidth]{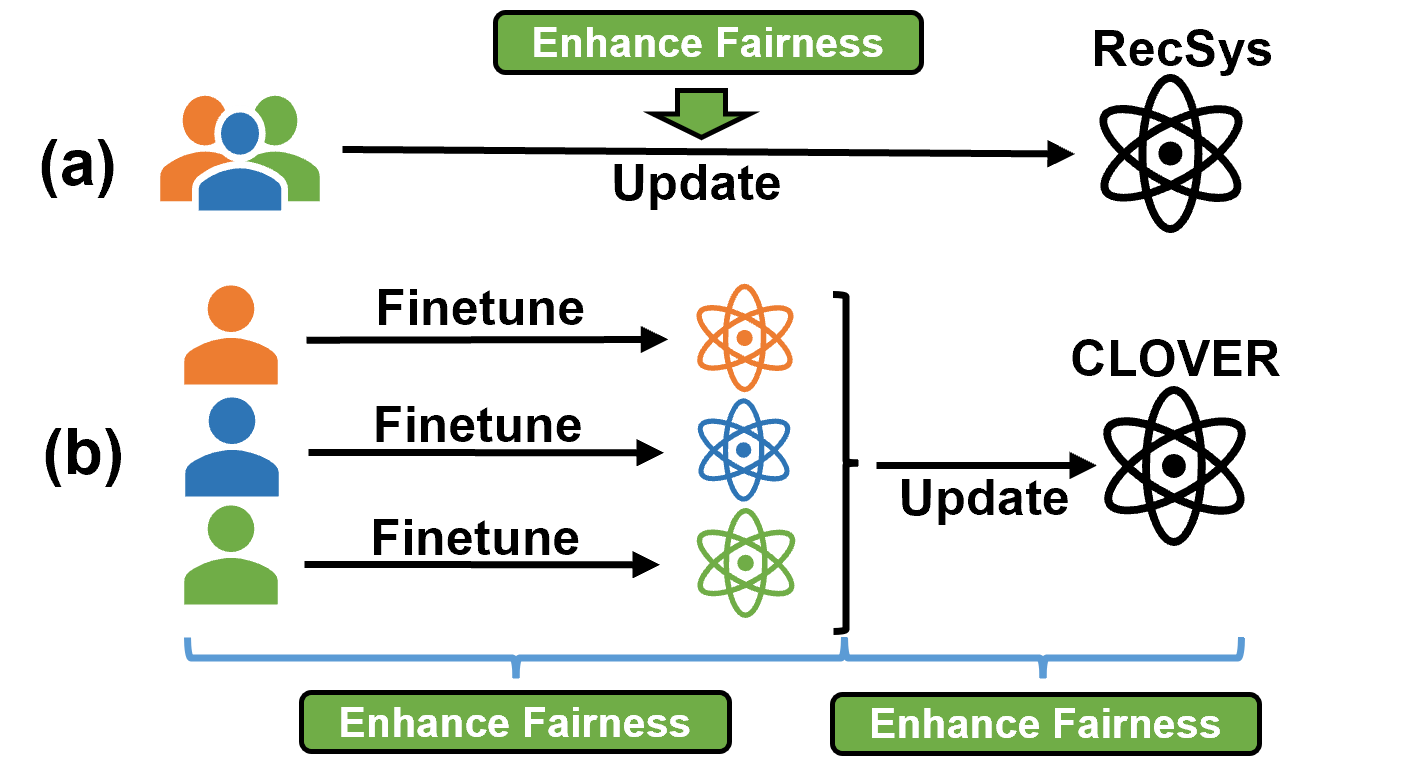}
  \caption{Difference between cold-start fair recommender system and existing fair recommender system.}
  \label{fig:diff}
  \vspace{-0.6cm}
\end{figure}

Recently, there has been growing attention on fairness considerations in the recommender systems \cite{beutel2019fairness, beigi2020privacy, wu2021learning, ge2021towards, li2021user}. The underlying unfairness issue in recommendation will hurt users’ or providers’ satisfaction as well as their interests in the platforms. Therefore, it is crucial to address the potential unfairness problems. The fairness measures in the recommender systems can be classified as follows. (1) \textbf{Individual fairness} refers to the protection of the sensitive attributes in the user modeling process against adversary attackers. There have been works \cite{beigi2020privacy, wu2021learning} relying on adversarial learning techniques \cite{goodfellow2014generative} that aim to match the conditional distribution of outputs given each sensitive attribute value. (2) \textbf{Counterfactual fairness} requires the recommendation results for a user to be unchanged in the counterfactual world where the user’s features remain the same except for certain sensitive features specified by the user. \citet{li2021towards} first explored this concept by generating feature-independent user embeddings to satisfy the counterfactual fairness requirements in the recommendation. (3) \textbf{Group fairness} aims to make the recommender systems not favor a particular demographic group over others. This fairness requirement is usually achieved by specific fairness regularization terms \cite{yao2017beyond, zhu2018fairness}. However, these methods are not designed to address the cold-start problem, and may render sub-optimal performance if used in this scenario. Furthermore, these methods only consider one specific kind of fairness. In other words, the effects and connections among different kinds of fairness in the recommender systems have not been studied.

To address these limitations, in this paper, we propose a novel comprehensive fair meta-learned recommender framework, {\it \model}, for ensuring the fairness of cold-start recommendation models. It offers a general training paradigm that is applicable to any meta-learned recommender system. We formulate various fairness constraints as the adversarial learning problem with two main components: the unfairness discriminator that seeks to infer users’ sensitive-attribute information, and the fair results generator that prevents the discriminator from inferring the sensitive information. We systematically illustrate the three kinds of fairness - individual fairness, counterfactual fairness, and group fairness, in the recommender system, and improve them in a unified manner via the multi-task adversarial learning scheme. Figure \ref{fig:diff} shows the difference between the existing fair recommender systems and the meta-learned fair recommender systems with cold-start, where we need to consider imposing fairness in both the inner loop and outer loop. To summarize, the major contributions of this paper are outlined as follows:
\begin{itemize}[leftmargin=*]
    \item We systematically illustrate the comprehensive fairness issues in the recommender systems and formulate enhancing comprehensive fairness as the multi-task adversarial learning problem.
    \item We propose \model, which is carefully designed to impose fairness in the framework of meta-learning with the interleaving training procedure. To the best of our knowledge, we are the first to explore the fair meta-learned recommender system.
    \item We demonstrate \model~ on the representative meta-learned user preference estimator on three real-world data sets. Empirical results show that \model~ achieves comprehensive fairness without deteriorating the overall cold-start recommendation performance.
\end{itemize}
The rest of the paper is organized as follows. We show the preliminary definition in Section \ref{section:preliminary} and introduce the proposed \model~ in Section \ref{section:method}. Then we present the experimental results in Section \ref{section:experiments}. Section \ref{section:related work} briefly discusses the existing work. In the end, we conclude the paper in Section \ref{section:conclusion}.

%% file: 3-methodology.tex
\section{Preliminary}
\label{section:preliminary}
In this section, we introduce the cold-start problem and the fairness considerations in the recommender systems. 
\subsection{Problem Statement}
\label{subsection:state}

Given a user $u$ with the profile $x_{u}$ and limited \textit{existing} rated items $I_{u}^e$, each item $i\in I_{u}^e$ is associated with a description profile $p_i$ and the corresponding rating $y_{ui}$. Our goal is to predict the rating $y_{ui^q}$ by user $u$ for the new \textit{query} item $i^q \in I_u^q$. Here, $I_u^q$ stands for the items needed to predict. There are two data sets for each user, one for fine-tuning, the other for testing. We define the existing fine-tuning data for each user $u$ as $D_u^e= \{x_u,p_i,y_{ui}\}_{i \in I_u^e}$, and query data as $D_u^q=\{x_u,p_i,y_{ui}\}_{i \in I_u^q}$. Following the previous works \cite{lee2019melu,dong2020mamo,wei2020fast}, we treat the fast adaption on each user $u$ as a task $t_{u}$:
\begin{equation}
    t_{u}: (\theta, D_{u}^e) \xrightarrow{} \theta_{u},
\end{equation}
where $\theta$ is the meta-model learned from the meta-learning recommender system, and $\theta_{u}$ is the personalized model fine-tuned on $D_u^e$ for each user. Cold-start recommendation focuses on new users arriving after the training stage. Let $U^f$ represent the sets of \textit{fresh} users that will arrive in the system. We aim to learn a proper $\theta$ on existing users that generalizes well on new users $U^f$, i.e., solving the fast adaption tasks.


\subsection{User-oriented Fairness}
\label{subsection:fairness}
Next, we give the comprehensive fairness definition of a recommender system. A fair recommender system should reflect the personalized preferences of users while being fair with respect to sensitive information. Specifically, the learned representation and recommender results should not expose the sensitive information that correlates with the users or show discrimination towards any individual or group of users. The fairness in the recommender system can be reflected from several different perspectives, which are summarized below.

\textbf{Individual Fairness} \cite{beigi2020privacy,wu2021learning}. Here, the fairness requirements refer to not exposing sensitive feature in the user modeling process against attackers. We define such fairness formally as:
\begin{equation}
    IF=\frac{1}{|U^f|}\max_g\sum_{u\in U^f} M(\hat{a}_u=g(e_u), a_u)
\end{equation}
where $e_u$ is the representation of user $u$, $a_u$ is the sensitive information of $u$, $g$ is the user representation attacker, aiming to predict the sensitive information from the user representation, and $M$ is the evaluation metric of the prediction performance. This definition requires the user modeling network to defend against any possible attacker that tries to hack the sensitive information. A lower $IF$ score indicates a fairer recommender.

\textbf{Counterfactual Fairness} \cite{li2021towards}. This is a causal-based fairness notion \cite{kusner2017counterfactual}. It explicitly requires that for any individual, the predicted result of the learning system should not discriminate towards the users' sensitive information. We will imagine a counterfactual world here in which we only make an intervention on the user’s sensitive information while making other independent features unchanged, and we expect the prediction to be the same as in the real world. We first refer to the definition of counterfactually fair recommendation in \cite{li2021towards}. 
\begin{definition}[Counterfactually fair recommendation \cite{li2021towards}]\label{counterfactual fairness}
A recommender model is counterfactually fair if for any possible user $u$ with features $X=x$ and $A=a$:
$$
P\left(L_{a} \mid X=x, A=a\right)=P\left(L_{a^{\prime}} \mid X=x
, A=a\right)
$$
for all $A$ and for any value $a^{\prime}$ attainable by $A$, where $L$ denotes the recommendation results for user $u$, $A$ is the user's sensitive feature and $X$ are the features that are independent on $A$.
\end{definition}
This definition requires that for any possible user, sensitive information $A$ should not be a cause of the recommendation results. In our setting of explicit rating prediction, we can formally define the counterfactual fairness as:
\begin{equation}
    CF=\frac{1}{|U^f|}\sum_{u\in U^f} |R(L_{a_u}^u \mid X=x, A=a_u)-R(L_{a^{\prime}_u}^u \mid X=x, A=a_u)|
\end{equation}
where $R$ is the performance of the recommendation, in our case the MAE (mean absolute error) of the rating prediction. $L_{a_u}^u$ is the recommendation results of user $u$ with sensitive information $a_u$. For simplicity, we abbreviate $R(L_{a_u}^u \mid X=x, A=a_u)$ into $R(u)$ in the following to represent the recommendation performance of user $u$. A lower $CF$ score indicates a fairer system.
\begin{figure}[t]
  \centering
  \includegraphics[width=1.0\linewidth]{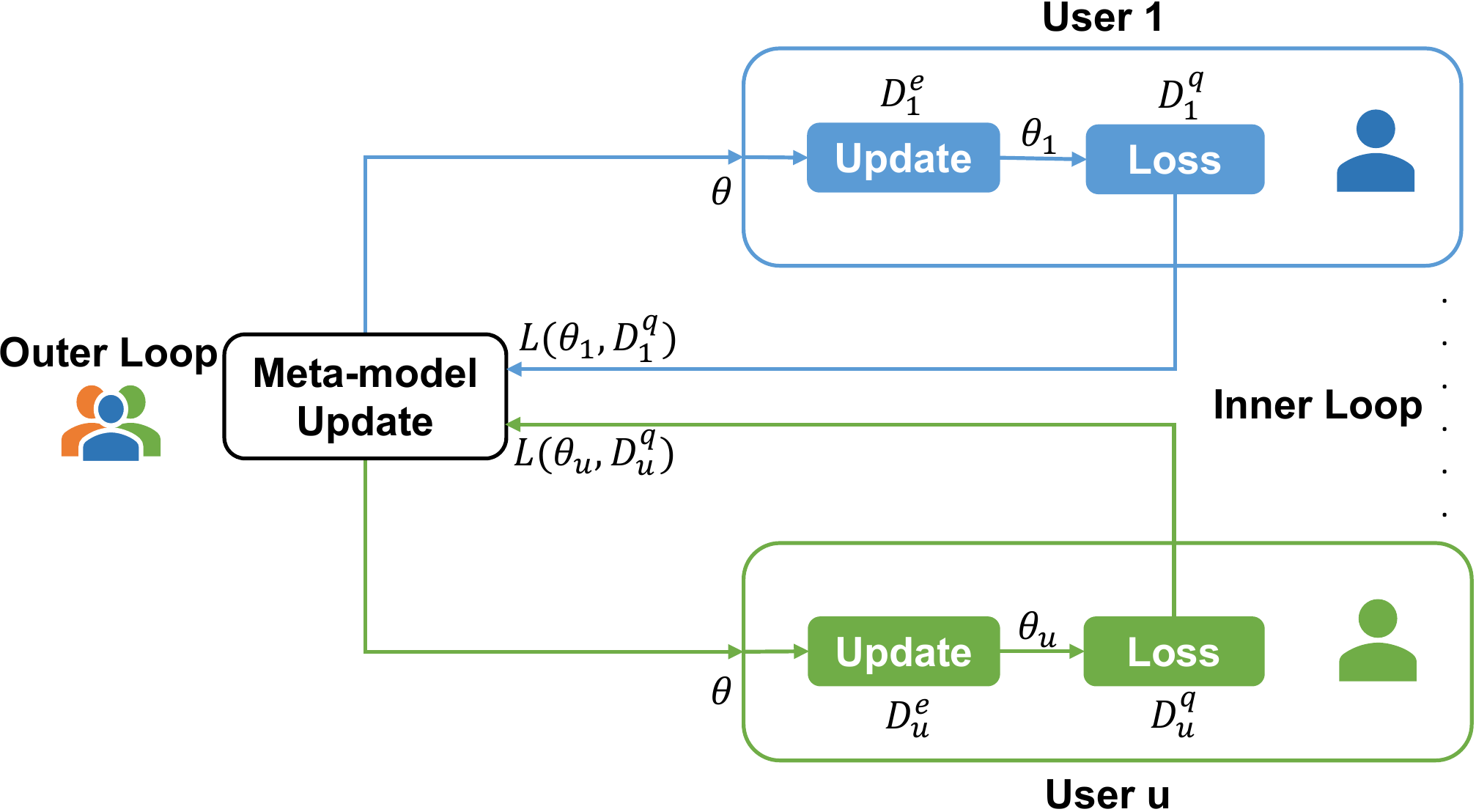}
  \caption{The training process of the meta-learned recommender system.}
  \label{fig:meta}
  \vspace{-0.4cm}
\end{figure}

\textbf{Group Fairness} \cite{liu2022dual,fu2020fairness,li2021user}. Group fairness requires the protected groups to be treated similarly as the advantaged group in the recommender system. The recommender systems will achieve fairness when users with two different attribute values maintain the same recommendation performance. We consider grouping testing users as $A_1$ and $A_2$ based on their sensitive information. More formally, the group recommendation unfairness \cite{fu2020fairness} is defined as follows:
\begin{equation}
    GF=|\frac{1}{|A_1|}\sum_{u_1\in A_1} R(u_1) - \frac{1}{|A_2|}\sum_{u_2\in A_2} R(u_2)|
\end{equation}
where notation $R$ is a metric that evaluates the recommendation performance, $|A_1|$ is the number of users in group $A_1$. Lower GF also represents better fairness performance.


These are all the desired properties of a fair recommender system. We regard the combination of these three requirements as comprehensive fairness and aim to design algorithms to improve comprehensive fairness simultaneously. In this work, we propose to design the comprehensive framework \model~ to incorporate all these fairness requirements. For counterfactual fairness and group fairness, our definition here mainly focuses on the case of binary sensitive attributes. For sensitive attributes with multiple values, we can measure the fairness with the sum of the differences between all possible other values of the sensitive attribute. In addition, exploring the fairness issues within the combination of multiple sensitive attributes is left as the future work.


\section{Proposed \model\ Framework}
\label{section:method}
In this section, we first briefly describe the training process of the meta-learned recommender system. Next, we introduce how to mitigate the comprehensive fairness issues in the recommender system by designing a novel multi-task adversarial learning component. Then, \model~ is proposed to build a fair recommender system in the cold-start meta-learning scenario. Finally, we demonstrate how to instantiate our generic method on the representative meta-learned user preference estimator. Note that our framework can also be applied to other methods, which we'll show in the experiments.

\subsection{Meta-learned Recommender System}
\label{subsection:meta}
The core idea of the meta-learned recommender system is learning to solve the fast adaption task $t_u$ on new users. We follow the most representative meta-learning recommender framework \cite{lee2019melu,dong2020mamo,wei2020fast} to show the workflow. To learn the parameters of a new task, the meta-model aims to provide a promising initialization by learning from various similar tasks. Then from the learned parameters, the meta-model can be further fine-tuned on the new task (user) with limited interactions to achieve personalized recommendation. The workflow of the meta-learned recommender system is shown in Figure \ref{fig:meta}. The framework iteratively updates the meta-model according to the following procedure:
\begin{itemize}[leftmargin=*]
    \item \textbf{Inner loop.} For each task $t_u$, the framework initializes $\theta_u$ with the latest parameters of the meta-model $\theta$, and updates $\theta_u$ according to the user's training data.
    \item \textbf{Outer loop.} The framework updates the meta-model $\theta$ by minimizing the recommendation loss of $\theta_u$ regarding each user $u$ to provide a promising initialization for every user.
    \end{itemize}

In the inner loop, the model parameters regarding a user $u$ will be updated iteratively as follows:
\begin{equation}
\begin{split}
    \theta_u=\theta_u-\alpha\nabla_{\theta_u} L(f_{\theta_u}, D_u^e),\\
\end{split}
\vspace{-0.2cm}
\end{equation}
where $\alpha$ is the learning rate of user parameter update, $L(f_{\theta_u}, D_u^e)$ denotes the recommendation loss (e.g., the cross-entropy log loss \cite{berg2017graph}) on data $D_u^e$. $f_{\theta_u}$ suggests the loss is parametrized by parameter $\theta_u$. The model parameter $\theta_u$ of user $u$ is initialized by the  meta-model parameter $\theta$. $D_u^e$ is the user's existing training data, as introduced in Section~\ref{subsection:state}.

In the outer loop, the meta-model parameters will be updated by summing up all user $u$'s specific loss $L(\theta_u, D_u^q)$ together to provide a promising initialization. $D^q_{u}$ is the query data set according to Section~\ref{subsection:state}. Specifically, in each step of training, the parameters are updated as follows:
\begin{equation}
    \theta= \theta-\beta \nabla_{\theta} \sum_{u\in B}L(f_{\theta_{u}},D_{u}^q)
\end{equation}
where $B$ is a set of users involved in the batch, $\beta$ is the learning rate of the meta-model parameters. In the following section, we abbreviate the notation $D$ in the above loss function for simplicity.

After introducing the training process of the meta-model. We will then discuss the testing process. For evaluation of new user $u^f\in U^f$, the framework will initialize the user model parameters with the meta-model $\theta$, and the user model is then fine-tuned with the user's observed interaction data $D_{u^f}^e$. Finally, the fine-tuned model will be applied to make recommendations.

\subsection{Comprehensive Unfairness Mitigation}
Next, we will discuss how to comprehensively mitigate the fairness issues in the recommender systems. The naive solution is to add fairness regularization into the training process of the meta-learned recommender system. However, this approach faces the following flaws. 
\begin{itemize}[leftmargin=*]
    \item In the inner loop update, the meta-model will be fine-tuned only according to the users' data. In this case, we are unable to calculate the fair regularization loss as it requires information from different groups of users. Thus, this kind of method is not appropriate for the meta-learned recommender system. 
    \item A strong fair recommender system should not expose the sensitive information that correlates with the users in any situation. The unknown attackers are not explicitly considered in the regularization-based approaches and therefore the performance may vary; whereas adversarial learning methods train the recommender to defend against any possible model that tries to hack the sensitive information.
\end{itemize}
In this paper, we formulate comprehensive fairness as the multi-task adversarial learning problem, where different fairness requirements are associated with different adversarial learning objectives. Our basic idea is to train a recommender and the adversarial attacker jointly. The attacker seeks to optimize its model to predict the sensitive information. The recommender aims to extract users' actual preferences and fool the adversary in the meantime. In this way, the sensitive information will be removed and the recommender is encouraged to provide fair results. We model this as a min-max game between the recommender and the attack discriminator. We argue the different fairness requirements in \ref{subsection:fairness} can be represented as yielding identical distribution concerning the sensitive attribute at either representation or prediction level. As shown in the Proposition 2 of \cite{goodfellow2014generative}, if the generator $G$ and the discriminator $D$ have enough capacity, and at each iteration, the discriminator $D$ is allowed to reach its optimum and the generator $G$ is updated to fool the discriminator, then the generated data distribution will converge to the distribution of the real data. Therefore, in our model with the sensitive information discriminator, upon convergence, the distribution of generated recommender representation and prediction with distinct sensitive attributes will be the same. Next, we will detail how to implement it in the recommender system.

Specifically, the model parameters consist of two parts $\theta=\{\theta^r, \theta^d\}$, where $\theta^r$ is the parameter of the recommender model and $\theta^d$ represents the sensitive information discriminator parameter. The objective function can be written as follows:
\begin{equation}
    \min_R \max_D L=L(f_{\theta^r, \theta^d})=l_R(f_{\theta^r})-l_D(f_{\theta^r, \theta^d})
    \vspace{-0.2cm}
\end{equation}
\begin{equation}
    l_D(f_{\theta^r, \theta^d}))=l_D(a_{u},\hat{a}_{u})=-\frac{1}{N}\sum_{u=1}^{N}\sum_{c=1}^Ca_{u}^c log(\hat{a}_{u}^c)
\end{equation}
where $L$ consists of the recommender loss $l_R$ and the sensitive attribute discriminator loss $l_D$. $L(f_{\theta^r, \theta^d})$ implies the loss is parameterized by the $\theta^r, \theta^d$. Note that the recommender loss $l_R$ is unrelated with the discriminator parameter $\theta^d$. $C$ is the number of classes of the sensitive attribute, $\hat{a}_{u}^c$ is the probability that the sensitive attribute of user $u$ is predicted by the discriminator to be class $c$, $\hat{a}_{u}$ is the predicted probability vector for user $u$, and $a_u$ is the one-hot vector of the sensitive information in which 1 denotes the ground-truth class. The recommender is asked to minimize the recommendation error while fooling the discriminator to generate fair results. Meanwhile, the discriminator is optimized to predict the sensitive information from the user as accurately as possible. In this work, we choose to optimize the recommender and discriminator simultaneously as it performs the best. The next problem is how to relate the adversarial learning problem with the comprehensive fairness mitigation. We argue that we can achieve comprehensive fairness by conducting both the representation and prediction level adversarial learning.

\begin{figure}[t] 
\centering
\includegraphics[width=0.95\linewidth]{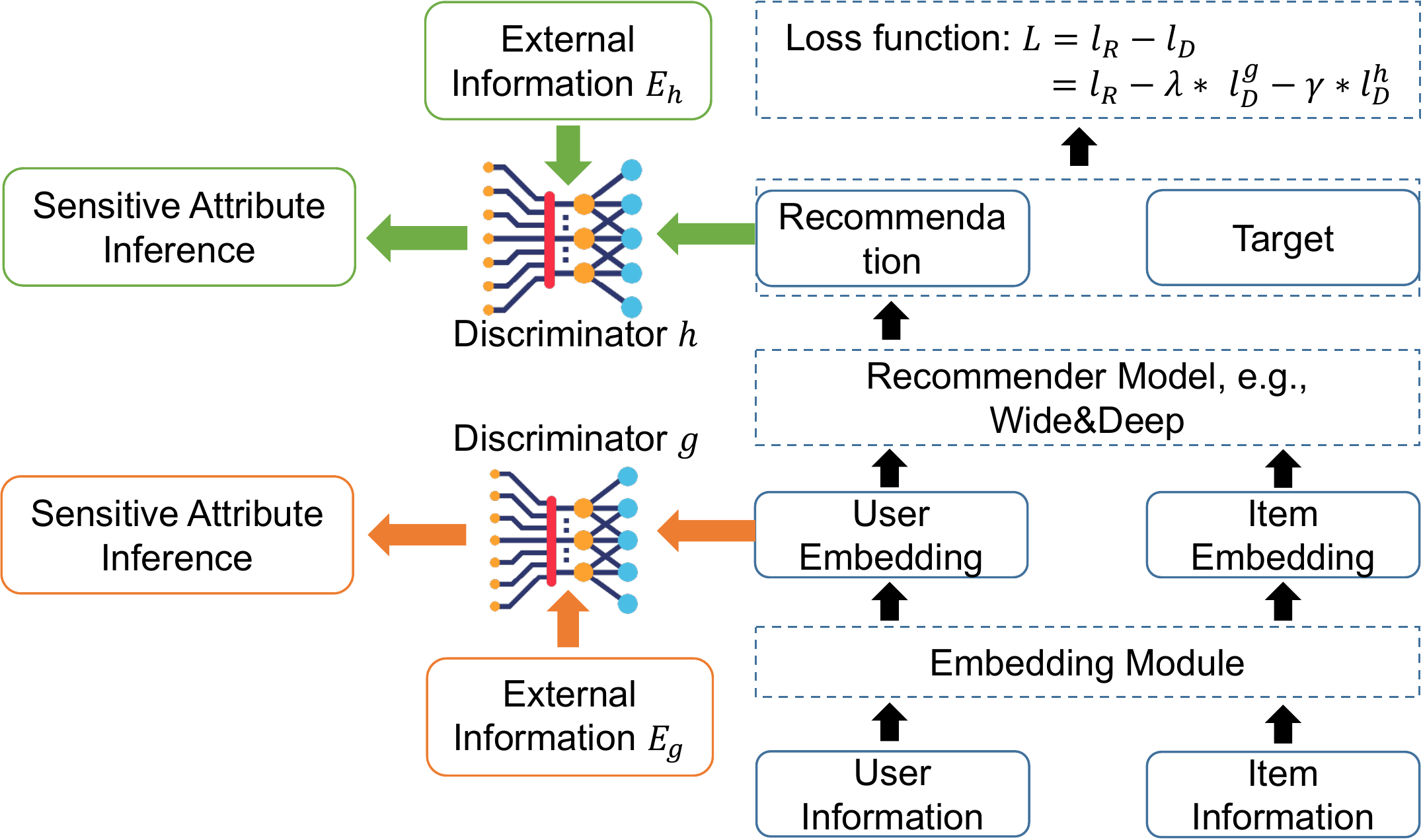}
\caption{The network structure of proposed \model.}
\label{fig:structure}
\vspace{-0.4cm}
\end{figure}

For the \textbf{individual fairness}, from the definition in Subsection \ref{subsection:fairness}, we can find that it requires the user modeling process of the recommender system to protect against the most powerful hacker from inferring the sensitive information. Therefore, we conduct the representation level adversarial learning which aims to generate the user embedding irrelevant to the sensitive information. More specifically, for a user $u$, the discriminator will predict the sensitive information according to the corresponding user representation:
\begin{equation}
    l_D^g = l_D(a_u, \hat{a}_u=g(e_u, E_g))
    \vspace{-0.1cm}
\end{equation}
where $g$ is the representation discriminator, $e_u$ is the user embedding of $u$. Different from the traditional adversarial game, we also input the external information $E_g$ for better performance. Here we concatenate the corresponding historical rating $y_{ui}$ with the user embedding as the input information. The benefits are two-fold: (1) Due to the external information $y_{ui}$ that the discriminator has access to, the recommender cannot hope to fully fool the discriminator, since the external rating information can give some insights of the users' sensitive information. In this way, the recommendation generator will focus more on removing the sensitive information of users. (2) Adding such external information requires the learned representation to be independent of the sensitive attributes conditioned on any values of $y$. It sets higher requirements for model learning and will be beneficial for comprehensive fairness, which we will show in the experiments section.

As for the \textbf{counterfactual fairness}, we prove it can be achieved with the above representation level adversarial learning for individual fairness. The proof is in the following.
\begin{proposition}
If the adversarial game for individual fairness converges to the optimal solution, then the rating prediction recommender which leverages these representations will also satisfy the counterfactual fairness.
\end{proposition}
\textit{Proof.} The downstream recommender uses the representation $e_u$ for predicting the rating $\hat{y}_{ui}$ of user $u$, thus forming a Markov chain $a_u \rightarrow e_u \rightarrow \hat{y}_{ui}$ (user sensitive attribute $a_u$ won't affect the recommender prediction via item embedding $e_i$). As we discussed before in this section, if the adversarial learning converges to the optimal, the generated recommender results will be independent of the sensitive attribute, i.e., the mutual information between the sensitive attribute $a_u$ and the representation $e_u$ for any given user $u$ is zero: $I(a_u;e_u) = 0$. Using the properties of inequality and non-negativity of mutual information:
\begin{equation}
    0 \leq	 I(a_u; \hat{y}_{ui}) \leq I(a_u; e_u) \And I(a_u; e_u)=0 \Longrightarrow I(a_u; \hat{y}_{ui})=0
\end{equation}
Therefore, the prediction $\hat{y}_{ui}$ for any given user $u$ is independent of the sensitive attribute $a_u$, and consequently, the downstream recommender satisfies the counterfactual fairness.$\blacksquare$

Concerning the \textbf{group fairness}, it requires the recommendation performance of users to be identical between different groups. Here we consider the mean absolute error $MAE=|y_{ui}-\hat{y}_{ui}|$. The goal can then be interpreted as to achieve the same $\hat{y}_{ui}$ across groups given true rating value $y_{ui}$, which will be satisfied when the discriminator cannot predict the sensitive attribute with a high accuracy given $y_{ui}$ and $\hat{y}_{ui}$. Thus we concatenate the predicted logits $\hat{y}_{ui}$ and $y_{ui}$ as the input to carry out the prediction-level adversary game:
\begin{equation}
    l_D^h = l_D(a_u, \hat{a}_u=h(y_{ui}, \hat{y}_{ui}, E_h))
    \vspace{-0.1cm}
\end{equation}
where $h$ is the prediction discriminator, $E_h$ is the external information for $h$. Here we regard the item embedding $e_i$ as the additional information. Except for the reasons mentioned above, the item data usually contain some sensitive information. For example, males tend to borrow some science fictions, while females may prefer the romance novels. In this way, we can remove the encoded bias in the item data.

To comprehensively mitigate the fairness issues, we perform multi-task learning with both representation and prediction level adversarial learning as follows:
\begin{equation}
    l_D = \lambda*l_D^g+\gamma*l_D^h
\end{equation}
where the hyper-parameters $\lambda$ and $\gamma$ are used to balance the contribution of these two losses. As $l_D$ will be further combined with recommendation loss $l_R$ for training, we use two hyper-parameters here to control the loss function more flexibly. The whole network structure is shown in the figure \ref{fig:structure}.


\subsection{Fair Meta-Learned Recommender System}
In the following, we will discuss how to achieve fairness in the context of the meta-learned recommender system. As shown in Subsection \ref{subsection:meta}, tackling the fairness problem in such a setting is challenging, since it is trained with a bi-level interleaving learning procedure, comprised of the inner loop update that fine-tunes on each user and outer loop update that produces the initialization of all users. Therefore, it remains elusive when and how fairness should be enhanced to strike a graceful balance between fairness, recommendation performance, and computational efficiency. About when to update, we consider the fairness enhancement from both the inner and outer loops perspectives. As for how to update, we denote the recommender task as $T_r$, and disentangle the adversarial game into two objectives $T_1$ and $T_2$. The three tasks can be shown as follows:
\begin{itemize}[leftmargin=*]
    \item $T_r$: the task of recommender loss minimization for $\theta^r$, which is required for all models.
    \item $T_1$: the task of optimizing the discriminator $\theta^d$ to predict the sensitive information.
    \item $T_2$: the task of updating the recommender $\theta^r$ to generate fair results by fooling the discriminator.
\end{itemize}
Then we will show when and how to update the meta-learned recommender system toward comprehensive fairness.

In the \textbf{outer loop}, optimizing the loss on query data represents better performance on testing data after fine-tuning. We hope the meta-model can provide a general initialization such that the fine-tuned model of various users will yield fair results. Therefore, we need to consider both the $T_1$ and $T_2$ tasks. The optimization process of the objective can be formulated as:
\begin{equation}
    \theta^r= \theta^r-\beta \nabla_{\theta^r} \sum_{u\in B}L(f_{\theta_{u}})
\vspace{-0.1cm}
\end{equation}
\begin{equation}
    \theta^d= \theta^d+\beta \nabla_{\theta^d} \sum_{u\in B}L(f_{\theta_{u}})
\end{equation}
where $\theta_u=\{\theta_u^r,\theta_u^d\}$ is initialized by $\theta=\{\theta^r,\theta^d\}$.

\begin{algorithm}[t]
    \caption{\model: Comprehensive Fair Meta-learned Recommender System}
    \label{alg:fairmeta}
    \begin{flushleft}
    \textbf{Input:} Traing user distribution:$d(u)$; $\alpha, \beta$: learning rate hyper-parameters; $\lambda$, $\gamma$: adversarial learning trade-off. \\
    \end{flushleft}
    \begin{algorithmic}[1]
    
    \STATE /* Training on the existing users */
    \STATE Randomly initialize $\theta$, where $\theta=\{\theta^r,\theta^d\}$;
    \STATE Denote $L=l_R-l_D=l_R-\lambda*l^g_D-\gamma*l^h_D$
    \WHILE{not converge}
        \STATE Sample batch of users $B\sim d(u)$;
        \FOR{user $u$ in $B$}
            \STATE Set $\theta^r_u,\theta^d_u = \theta^r,\theta^d$; $\theta_u=\{\theta^r_u,\theta^d_u\}$;
            \STATE 
                Inner loop update:
                \begin{equation*}
                \begin{split}
                &\theta^r_u \gets \theta^r_u - \alpha \nabla_{\theta^r_u} l_R(f_{\theta_u^r}) \\
                &\theta^d_u \gets \theta^d_u + \alpha \nabla_{\theta^d_u} L(f_{\theta_u})\\
                \end{split}
                \end{equation*}
            
            
        \ENDFOR
        \STATE 
        Outer loop update:
        \begin{equation*}
            \begin{split}
                & \theta^r \gets \theta^r - \beta \sum_{u \in B} \nabla_{\theta^r} L(f_{\theta_u}) \\
                & \theta^d \gets \theta^d + \beta \sum_{u \in B} \nabla_{\theta^d} L(f_{\theta_u}) \\
            \end{split}
        \end{equation*}
    \ENDWHILE
    \STATE /* Testing on the new users */
    \FOR{user ${u^f}$ in $U^f$}
    \STATE 
                Finetune:
                \begin{equation*}
                \begin{split}
                &\theta^r_{u^f} \gets \theta^r_{u^f} - \alpha \nabla_{\theta^r_{u^f}} l_R(f_{\theta_{u^f}^r}) \\
                \end{split}
                \end{equation*}
    \STATE Perform recommendation for user $u^f$ base on $\theta^r_{u^f}$
    \ENDFOR
    \end{algorithmic}
\end{algorithm}

\setlength{\textfloatsep}{0pt}

For the \textbf{inner loop}, a natural question is whether we still need to consider both of the tasks $T_1$ and $T_2$. First of all, we argue task $T_1$ is still essential. If the discriminator is not updated, it will be easy for the recommender to fool the discriminator. Then it will fail to achieve the purpose of guiding the meta-model to provide initialization that can produce fair results. Then for task $T_2$, we propose to remove it from the inner loop. There are three main reasons for this:
\vspace{-0.7cm}
\begin{itemize}[leftmargin=*]
    \item \textbf{Model stability.} For fast adaptation, the user model will be fine-tuned by only a few steps of gradient descent, whereas it usually takes a longer time for the adversarial game to reach the desired equilibrium \cite{karras2019style}. Moreover, generative adversarial network is notoriously difficult to tune \cite{goodfellow2014generative}. If we include the task $T_2$ in the inner loop, we will need to maintain two sets of adversarial networks, which will greatly increase the training instability of the model.
    
    \item \textbf{Training efficiency.} The bulk of computation is largely spent performing the adversarial game in the inner loop as it requires multiple times of gradient descent. Optimizing the task $T_2$ will additionally increases the cost of fine-tuning during both training and deployment. Thus, eliminating such an operation significantly accelerates the training time and improves efficiency. 
    \item \textbf{Privacy.} Task $T_2$ requires each user's sensitive information during fine-tuning. If we do not execute this task, we can perform fair recommendations without leveraging the sensitive information of new users during deployment, and thus protect the privacy of new users. In other words, we can train on risk-free, user-approved privacy data and then perform privacy-preserving fair recommendations for new users.
\end{itemize}
Due to the above reasons, we choose to perform only the optimization of task $T_1$ as follows:
\begin{equation}
    \theta^d_u= \theta^d_u+\beta \nabla_{\theta^d_u} L(f_{\theta_{u}})
\end{equation}
As task $T_2$ is not required, we just need to optimize the task $T_r$ for $\theta^r_u$:
\begin{equation}
    \theta^r_u= \theta^r_u-\beta \nabla_{\theta^r_u} l_R(f_{\theta_{u}})
\end{equation}

During testing, the model will be fine-tuned on the user's existing data with the initialization parameters of the meta-model. Since we no longer need to update the meta-model during deployment, we only need to perform task $T_r$. The complete algorithm is shown in Algorithm \ref{alg:fairmeta}.

\vspace{-0.3cm}
\subsection{Instantiation to MELU}
To demonstrate how our proposed framework works, we provide an implementation based on Meta-Learned User Preference Estimator (MELU) \cite{lee2019melu}, a representative embedding model for cold-start recommendation. It introduces the idea of meta-optimization mentioned above into the cold-start scenario and builds a powerful network for the recommendation. Although it has shown great capacity at the cold-start recommendation, it does not consider fairness issues. Our framework \model~ is demonstrated based on the recommender network and meta-optimization process of MELU. Note that our framework is model-agnostic and can be applied to other meta-learned cold-start models, which we'll show in the experiments.

In the following, we will briefly introduce the recommender network and the associated loss. For the user embedding process, MELU generates each content embedding and uses the concatenated embeddings. When the number of user contents is $P$, it defines the embedding vector $e_u$ for user $u$ as follows:
\begin{equation}
    e_u = W^T_U\left[ E_{U}^1x_{u}^1;\, \cdots; E_{U}^Px_{u}^P \right]^\intercal+b_U
    \label{eq:user_emb}
\end{equation}
where $x_{u}^p$ is a $d_p$-dimensional one-hot vector for categorical content $p$ of user profile $x_u$, and $E_{U}^p$ represents the $d_e\times d_p$ embedding matrix for the corresponding categorical content of the users. $d_e$ and $d_p$ are the embedding dimension and the number of categories for content $p$, respectively. In order to increase the representation capability of user embedding to meet the fairness requirements, a one-layer fully-connected neural network is added in the modeling process. The item embedding $e_i$ of item $i$ is obtained in the same way. Following \cite{lee2019melu}, the initial embedding tables $E$ are only updated in the outer loop to guarantee the stability of model training. Then a N-layer fully-connected neural network decision-making module $F_N$ is constructed to estimate the user preferences, e.g., ratings. The module can be formulated as:
\begin{equation}
    \hat{y}_{ui} = F_N([e_u; e_i])
    \label{eq:decisionlayer}
\end{equation}
where the input is the concatenation of user and item embeddings, and $\hat{y}_{ui}$ is user $u$’s estimated preference for item $i$. The above together form the recommendation model $\theta_r$. As for the recommendation loss, we adopt one of the most widely used loss functions, cross-entropy classification loss, as it can be applied to both the explicit rating prediction \cite{berg2017graph} and implicit click forecasting \cite{he2018nais}:
\begin{equation}
    l_R^{ui} = -\sum_{c=1}^{|Y|}y_{ui}^c log(\hat{y}_{ui}^c)
\vspace{-0.2cm}
\end{equation}
where $l_R^{ui}$ here is the loss for each user-item rating pair, the loss $l_R$ is calculated on all interaction pairs, $|Y|$ is the number of ratings, and $\hat{y}_{ui}^c$ is the probability that the predicted rating of user-item interaction pair $ui$ is $c$.

%% file: 4-experiments.tex
\section{Experiments}
\label{section:experiments}

In this section, we conduct experiments to evaluate the performance of our proposed \model. Our experiments intend to answer the following research questions:
\begin{itemize}[leftmargin=*]
    \item \textbf{RQ1: }Does \model~ enable the cold-start recommender system to learn fair and accurate recommendations?
    \item \textbf{RQ2: }How do different components in our framework contribute to the performance?
    \item \textbf{RQ3: }How do different hyper-parameter settings (e.g. $\lambda$, $\gamma$) affect the recommendation and fairness performance?
    \item \textbf{RQ4: }How is the generalization ability of our proposed \model~ on other cold-start meta-learning models?
\end{itemize}
\subsection{Experimental Settings}

\begin{table}[t]
\setlength{\abovecaptionskip}{0.cm}
\centering
\caption{Statistics of Datasets.}
\label{tab:data}
\scalebox{0.78}{
\begin{tabular}{c|c|c|c}
\toprule
    \textbf{Dataset}           & \textbf{ML-1M} & \textbf{BookCrossing} & \textbf{ML-100K} \\ 
    \hline
    No. of users     & 6,040              & 278,858 &  943 \\ 
    No. of items     & 3,706              & 271,379 &  1,682 \\ 
    No. of ratings   & 1,000,209          & 1,149,780 &  100,000 \\ 
    Sparsity            & 95.5316\%          & 99.9985\% &  93.6953\%\\ 
    \hline 
    User contents       & \makecell{Gender, Age, \\ Occupation,\\ Zip code}  & \makecell{Age, Location}  & \makecell{Gender, Age, \\ Occupation,\\ Zip code}   \\ 
    \hline 
    Item contents       & \makecell{Publication year, \\ Rate, Genre, \\ Director, Actor}  & \makecell{Publication year, \\ Author, Publisher } &  \makecell{Publication year, Genre} \\ 
    \hline 
    Range of ratings    & 1 $\sim$ 5         & 1 $\sim$ 10 & 1 $\sim$ 5 \\ 
\bottomrule
\end{tabular}}
\end{table}

\begin{table*}[htbp]
\setlength{\abovecaptionskip}{0.cm}
\caption{Experimental results on the three datasets averaged over five independent runs. Arrows ($\uparrow$, $\downarrow$) indicate the direction of better performance. \model~ keeps the predictive power of the original recommender model while improving their fairness. Bold values indicate the best performance with regard to the meta-learned recommender model.}
\label{table:main}
\scalebox{0.69}{
\begin{tabular}{c|ccccc|ccccc|ccccc}
\hline
\multirow{2}{*}{} & \multicolumn{5}{c|}{ML-1M}                                                                                                                & \multicolumn{5}{c|}{BookCrossing}                                                                                                         & \multicolumn{5}{c}{ML-100K}                                                                                                               \\ \cline{2-16} 
                  & \multicolumn{1}{c|}{MAE $\downarrow$}  & \multicolumn{1}{c|}{NDCG $\uparrow$} & \multicolumn{1}{c|}{AUC $\downarrow$}  & \multicolumn{1}{c|}{CF $\downarrow$}   & GF $\downarrow$                        & \multicolumn{1}{c|}{MAE $\downarrow$ }  & \multicolumn{1}{c|}{NDCG $\uparrow$} & \multicolumn{1}{c|}{AUC $\downarrow$ }  & \multicolumn{1}{c|}{CF $\downarrow$}   & GF $\downarrow$                         & \multicolumn{1}{c|}{MAE $\downarrow$ }  & \multicolumn{1}{c|}{NDCG $\uparrow$} & \multicolumn{1}{c|}{AUC $\downarrow$ }  & \multicolumn{1}{c|}{CF $\downarrow$ }   & GF $\downarrow$                         \\ \hline \hline
PPR               & 0.943\std{0.012}          & 0.672\std{0.004}          & 0.932\std{0.013}          & 0.187\std{0.058}          & 0.052\std{0.012}          & 2.374\std{0.015}          & 0.598\std{0.005}          & 0.854\std{0.051}          & 0.194\std{0.035}          & 0.101\std{0.013}          & 1.398\std{0.009}          & 0.486\std{0.005}          & 0.772\std{0.036}          & 0.095\std{0.032}          & 0.051\std{0.008}          \\ 
Wide\&Deep        & 0.815\std{0.007}          & 0.691\std{0.003}          & 0.945\std{0.024}          & 0.214\std{0.034}          & 0.055\std{0.014}          & 1.945\std{0.028}          & 0.627\std{0.008}          & 0.833\std{0.034}          & 0.156\std{0.029}          & 0.098\std{0.005}          & 1.256\std{0.002}          & 0.519\std{0.006}          & 0.863\std{0.044}          & 0.112\std{0.021}          & 0.046\std{0.012}          \\  \hline
DropoutNet        & 0.813\std{0.005}          & 0.702\std{0.004}          & 0.965\std{0.013}          & 0.204\std{0.021}          & 0.063\std{0.003}          & 1.855\std{0.002}          & 0.634\std{0.005}          & 0.889\std{0.051}          & 0.194\std{0.035}          & 0.103\std{0.011}          & 1.172\std{0.007}          & 0.544\std{0.005}          & 0.842\std{0.036}          & 0.131\std{0.018}          & 0.044\std{0.013}          \\ 
NLBA              & 0.795\std{0.006}          & 0.701\std{0.002}          & 0.971\std{0.024}          & 0.254\std{0.047}          & 0.056\std{0.007}          & 1.718\std{0.007}          & 0.651\std{0.008}          & 0.943\std{0.034}          & 0.156\std{0.029}          & 0.112\std{0.005}          & 1.213\std{0.003}          & 0.553\std{0.006}          & 0.891\std{0.044}          & 0.129\std{0.023}          & 0.047\std{0.004}          \\  \hline
MELU              & 0.743\std{0.008}          & 0.755\std{0.002}          & 1.000\std{0.000}          & 0.264\std{0.058}          & 0.054\std{0.011}          & \textbf{1.332\std{0.008}} & \textbf{0.723\std{0.006}} & 1.000\std{0.000}          & 0.259\std{0.094}          & 0.103\std{0.011}          & 0.892\std{0.009}          & 0.652\std{0.017}          & 0.974\std{0.016}          & 0.157\std{0.032}          & 0.048\std{0.014}          \\ 
MELU+BS                & 0.745\std{0.003}          & 0.744\std{0.003}          & 1.000\std{0.000}          & 0.277\std{0.047}          & 0.049\std{0.012}          & 1.343\std{0.005}          & 0.712\std{0.005}          & 1.000\std{0.000}          & 0.291\std{0.058}          & 0.101\std{0.010}          & 0.891\std{0.002}          & 0.651\std{0.011}          & 1.000\std{0.000}          & 0.174\std{0.019}          & 0.046\std{0.008}          \\ 
MELU+Reg               & 0.762\std{0.002}          & 0.748\std{0.002}          & 0.894\std{0.011}          & 0.145\std{0.033}          & 0.048\std{0.009}          & 1.357\std{0.008}          & 0.713\std{0.004}          & 0.912\std{0.023}          & 0.232\std{0.024}          & 0.097\std{0.011}          & 0.914\std{0.007}          & 0.648\std{0.013}          & 0.873\std{0.036}          & 0.132\std{0.012}          & 0.045\std{0.005}          \\ 
MELU+IPW               & 0.751\std{0.012}          & 0.751\std{0.002}          & 0.975\std{0.008}          & 0.187\std{0.013}          & 0.047\std{0.011}          & 1.365\std{0.004}          & 0.707\std{0.006}          & 1.000\std{0.000}          & 0.245\std{0.044}          & 0.096\std{0.007}          & 0.903\std{0.008}          & 0.643\std{0.005}          & 0.946\std{0.024}          & 0.117\std{0.038}          & 0.047\std{0.009}          \\ 
MELU+MACR              & 0.744\std{0.005}          & 0.750\std{0.004}          & 0.878\std{0.031}          & 0.116\std{0.027}          & 0.051\std{0.011}          & 1.352\std{0.006}          & 0.715\std{0.002}          & 0.867\std{0.016}          & 0.179\std{0.094}          & 0.097\std{0.013}          & 0.887\std{0.011}          & 0.656\std{0.002}          & 0.815\std{0.019}          & 0.108\std{0.016}          & 0.044\std{0.007}          \\ \hline
\textbf{CLOVER}            & \textbf{0.731\std{0.005}} & \textbf{0.756\std{0.005}} & \textbf{0.632\std{0.057}} & \textbf{0.044\std{0.014}} & \textbf{0.040\std{0.006}} & 1.352\std{0.006}          & 0.722\std{0.006}          & \textbf{0.546\std{0.032}} & \textbf{0.027\std{0.011}} & \textbf{0.089\std{0.009}} & \textbf{0.880\std{0.009}} & \textbf{0.666\std{0.003}} & \textbf{0.562\std{0.021}} & \textbf{0.032\std{0.015}} & \textbf{0.036\std{0.006}} \\ \hline
\end{tabular}}
\end{table*}

\paragraph{Evaluation}
To verify the effectiveness of our proposed method, we conduct experiments on three publicly accessible recommender system datasets: ML-1M, BookCrossing, and ML-100K. The statistics of the datasets are shown in Table~\ref{tab:data}. We provide the details of the datasets in Appendix~\ref{appendix:datasets}. To evaluate the performance of the user cold-start recommendation, we split the users into training, validation, testing datasets with a ratio of 7:1:2. Following exactly the settings in \cite{lee2019melu}, for each user $u$, to mimic the real-world situation, we use a limited, unfixed number of interactions of the user as the existing fine-tuned data $D_u^e$, and leave the last 10 interactions of the user as the query data $D_u^q$ for evaluation. For fair recommendation, we choose the gender attribute as the sensitive attribute for ML-100K and ML-1M datasets and regard the age attribute as the sensitive attribute for the BookCrossing dataset. For individual fairness, we aim to evaluate whether the sensitive information is exposed by the learned user representation. Similar to many works for fairness recommenders \cite{beigi2020privacy, wu2021learning}, we take the training users’ attributes as ground truth and train a linear classification model by taking the learned fair representations as input. We report the classification accuracy on the users in the test set for the individual fairness evaluation. We use the AUC metric to measure individual fairness performance. Concerning the counterfactual fairness, we perturb the sensitive attribute to measure whether it will affect the recommendation results. The results of all metrics in our experiments are averaged over all users. For metrics, we report MAE and NDCG to evaluate the rating prediction and ranking performance. We adopt AUC, CF (counterfactual), and GF (group) to show the performance of the three kinds of fairness. The function details of these metrics can be found in Appendix \ref{appendix:metrics}. For the baselines, we first compare with the content-based filtering methods PPR, Wide\&Deep and the traditional cold-start recommender system baselines DropoutNet and NLBA. For fairness consideration, due to the lack of relevant baselines on meta-learned recommender system, we test the following fairness approaches instead based on MELU: BS, Reg, IPW and MACR. The details of these baselines are shown in Appendix \ref{appendix: baselines}.

\subsection{Results (RQ1)}
Table \ref{table:main} shows the recommendation and fairness performance of all compared methods w.r.t. MAE, NDCG, AUC, CF, and GF on the ML-1M, BookCrossing, and ML-100K datasets under the new user cold-start settings. From the table, we have the following findings:
\begin{itemize}[leftmargin=*]
    \item We observe that MELU, where the meta-learning strategy is used for new users, has a much better performance than the DropoutNet, NLBA, and content filtering baselines. This demonstrates the advantage of leveraging meta-learned prior in the cold-start problem of recommender systems. 
    \item We notice the fairness performance of the MELU method is relatively worse than other cold-start and content-based baselines. This shows that while the meta-learned cold-start algorithm makes better use of user data and experience for the recommendation, it also increases the unfairness and bias in recommendation results. This also indicates the importance of designing a fair learning method in this setting.
    \item We find that \model~ substantially outperforms all baseline methods concerning the fairness performance while not sacrificing the recommendation performance. This indicates the success of the comprehensive adversarial learning framework. Meanwhile, in many cases, our proposed method can slightly improve the recommendation performance, which indicates that incorporating adversarial learning may regularize the recommender model and help model better convergence to optimal.
    \item As to the fairness baselines of MELU, we observe that they are mostly designed for group fairness, and cannot perform well on imposing individual and counterfactual fairness. Among the baselines, Reg and MACR perform the best. Reg explicitly adds the fairness regularization to the objective, and MACR incorporates the module to detect and remove the biased information in the learning procedure. However, they do not consider the characteristics and convergence of the meta-learned recommender system and fail to achieve the desired performance.
    \end{itemize}


\begin{table}[t]
\setlength{\abovecaptionskip}{0.cm}
\caption{Effect of different optimization objectives.}
\label{table:task}
\scalebox{0.85}{
\begin{tabular}{c|ccccc}
\hline 
              & MAE $\downarrow$           & NDCG $\uparrow$            & AUC $\downarrow$           & CF $\downarrow$            & GF $\downarrow$            \\ \hline
MELU          & 0.743\std{0.008}          & 0.755\std{0.002}          & 1.000\std{0.000}          & 0.264\std{0.058}          & 0.054\std{0.011}          \\
CLOVER w/o    & 0.739\std{0.004}          & 0.754\std{0.003}          & 1.000\std{0.000}          & 0.332\std{0.069}          & 0.056\std{0.013}          \\ 
CLOVER$_{T2}$     & 0.761\std{0.007}          & 0.741\std{0.004}          & 1.000\std{0.000}          & 0.247\std{0.042}          & 0.062\std{0.007}          \\ 
CLOVER$_{T1\&T2}$ & 0.749\std{0.005}          & 0.751\std{0.008}          & 0.792\std{0.035}          & 0.105\std{0.015}          & 0.051\std{0.008}          \\ \hline
CLOVER        & \textbf{0.731\std{0.005}} & \textbf{0.756\std{0.005}} & \textbf{0.632\std{0.057}} & \textbf{0.044\std{0.014}} & \textbf{0.040\std{0.006}} \\ \hline
\end{tabular}}
\end{table}

\subsection{Case Study}
\paragraph{Impact of different components (RQ2)}
In this part, we investigate the effectiveness of the proposed \model~by evaluating the impact of different components. First, we show how do the defined tasks $T_1$ and $T_2$ affect the performance of the model. Here we conduct ablation studies on \model~ on the ML-1M dataset to show the performance of different combinations of the tasks. Note that these experiments are all about the optimization of inner-loop. For outer-loop, we'll perform both $T_1$ and $T_2$ tasks. Specifically, we compare \model~ with its three variants: \model$_{w/o}$, \model$_{T_2}$, \model$_{T_1\&T_2}$. \model$_{T_2}$ ($_{T_1\&T_2}$) denotes that we optimize the task $T_2$ (${T_1\&T_2}$) in the inner loop and \model$_{w/o}$ means that we do not optimize either of the two tasks. In our framework~\model, we only incorporate the optimization of task $T_1$ in the inner loop. The results are shown in Table \ref{table:task}. From the table, we can find that removing task $T_1$ will greatly hurt the performance of recommendation, which validates the significance of optimizing the discriminator in the inner loop so as to generate fair recommendation results. Moreover, compared with the results that incorporate task $T_2$, we can observe its disadvantage for the model stability and will hurt the model performance.

Next, we explore the impact of the designed loss function. We compare \model~ with its four special cases: \model$_{w/o~E^h}$ ($_{w/o~E^g}$), where the external information $E^h$ ($E^g$) is removed from the input to the loss; \model$_{w/o~l_D^h}$ ($_{w/o~l_D^g}$), where we remove the adversarial loss function designed for the individual and counterfactual fairness. From Table \ref{table:loss}, we can have the following findings: each designed loss function can improve the corresponding fairness metrics, and combining the two loss functions to perform multi-task learning can further enhance the results. Furthermore, compared with removing the external information, \model~ performs better concerning all fairness metrics. We can also find adding external information $E^h$ not only improve the individual and counterfactual fairness but also substantially increases the group fairness metric, which verifies our assumption that it plays an important function in imposing comprehensive fairness.
\paragraph{Effect of Hyper-parameters (RQ3)} We find that our framework can improve the recommendation and fairness performance in a wide range of hyper-parameters. Due to the space limit, we provide the detailed results and analysis of parameter sensitivity of $\lambda$, $\gamma$ in Appendix~\ref{appendix: hyper-parameters}.

\begin{table}[t]
\setlength{\abovecaptionskip}{0.cm}
\caption{Effect of different adversarial loss functions.}
\label{table:loss}
\scalebox{0.8}{
\begin{tabular}{c|ccccc}
\hline
                   & MAE $\downarrow$          & NDCG $\uparrow$           & AUC $\downarrow$          & CF $\downarrow$           & GF $\downarrow$           \\ \hline
MELU               & 0.743\std{0.008}          & 0.755\std{0.002}          & 1.000\std{0.000}          & 0.264\std{0.058}          & 0.054\std{0.011}          \\
CLOVER w/o $E^h$   & 0.735\std{0.005}          & \textbf{0.758\std{0.002}}          & 0.667\std{0.041}          & 0.051\std{0.009}          & 0.044\std{0.006}          \\
CLOVER w/o $l_D^h$ & \textbf{0.730\std{0.004}}          & 0.754\std{0.006}          & 0.687\std{0.062}          & 0.067\std{0.013}          & 0.055\std{0.010}          \\
CLOVER w/o $E^g$   & 0.736\std{0.006}          & 0.752\std{0.003}          & 0.725\std{0.037}          & 0.056\std{0.015}          & 0.049\std{0.012}          \\
CLOVER w/o $l_D^g$ & 0.738\std{0.005}          & 0.754\std{0.008}          & 0.914\std{0.018}          & 0.198\std{0.021}          & 0.043\std{0.008}          \\ \hline
CLOVER             & 0.731\std{0.005} & 0.756\std{0.005} & \textbf{0.632\std{0.057}} & \textbf{0.044\std{0.014}} & \textbf{0.040\std{0.006}} \\ \hline
\end{tabular}}
\end{table}

\paragraph{Ability of Generalization (RQ4)}
To verify the generalization ability of our model, we test our framework on two other representative meta-learned cold-start recommender systems. MetaCS \cite{bharadhwaj2019meta} follows a similar idea of MeLU when constructing the recommender model while using a more flexible meta-update strategy to learn the model parameters. MAMO \cite{dong2020mamo} designed two memory matrices that can store task-specific memories and feature-specific memories to achieve personalized initialization and adaptation on the new users. We employ our \model~ on these two methods to show the effectiveness, where \model$_{MetaCS}$ (\model$_{MAMO}$) represents the method with \model~ framework applied on MetaCS (MAMO). The results are presented in Table \ref{table:general}. It can be seen that our \model~ brings consistent improvements that are model-agnostic, which clearly show the generalization ability of our framework.

%% file: 2-relatedwork.tex
\section{related work}
\label{section:related work}
In this section, we briefly review the related work on cold-start in recommendation, fairness in recommendation, and fairness in meta-learning.

\vspace{-0.1cm}
\subsection{Cold-start in Recommendation}
Cold-start is a common problem that recommender systems will face when there is insufficient history information in recommending items to users. Over the past years, meta-learning, also known as learning to learn, has been massively adopted to deal with the cold-start problem in the recommender system with great success. It enables models to quickly learn a new task with scarce labeled data by utilizing prior knowledge learned from previous tasks. \citet{lee2019melu} proposed MELU to learn the initial weights of the neural networks for cold-start users based on MAML \cite{finn2017model}. \citet{bharadhwaj2019meta} follows a similar idea of MELU when constructing the recommender model while using a more flexible meta-update strategy to learn the model parameters. \citet{dong2020mamo} designs two memory matrices that store task-specific memories and feature-specific memories to achieve personalized parameter initialization and adaptation for each user. \citet{lu2020meta} took the advantage of the heterogeneous information network and proposed semantic-specific meta-learning to address the cold-start problem. \citet{wei2020fast} equips meta-learning with collaborative filtering and dynamic subgraph sampling, which can leverage other users' history information to predict the user preference more accurately.  There're also many works that leverage meta-learning in different settings, such as online recommender \cite{du2019sequential, sun2021form}, session recommender \cite{song2021cbml}, bandit \cite{DBLP:journals/corr/abs-2201-13395} and sequential recommender \cite{wang2021sequential}, etc. However, the understanding and mitigation of fairness under the meta-learning framework is largely under-explored. In this work, we explore mitigating the fairness issues of the cold start meta-learned recommender system.
\vspace{-0.3cm}

\begin{table}[t]
\setlength{\abovecaptionskip}{0.cm}
\caption{Generalization ability of \model.}
\label{table:general}
\scalebox{0.8}{
\begin{tabular}{c|ccccc}
\hline
              & MAE $\downarrow$          & NDCG $\uparrow$           & AUC $\downarrow$          & CF $\downarrow$           & GF $\downarrow$           \\ \hline
MELU          & 0.743\std{0.008}          & 0.755\std{0.002}          & 1.000\std{0.000}          & 0.264\std{0.058}          & 0.054\std{0.011}          \\
CLOVER$_{MELU}$   & \textbf{0.731\std{0.005}} & \textbf{0.756\std{0.005}} & \textbf{0.632\std{0.057}} & \textbf{0.044\std{0.014}} & \textbf{0.040\std{0.006}} \\ \hline
MetaCS        & 0.721\std{0.005}          & \textbf{0.776\std{0.003}} & 1.000\std{0.000}          & 0.245\std{0.058}          & 0.061\std{0.015}          \\
CLOVER$_{MetaCS}$ & \textbf{0.720\std{0.002}} & 0.775\std{0.005}          & \textbf{0.578\std{0.037}} & \textbf{0.029\std{0.009}} & \textbf{0.044\std{0.008}} \\ \hline
MAMO          & 0.717\std{0.003}          & 0.781\std{0.004}          & 1.000\std{0.000}          & 0.331\std{0.058}          & 0.057\std{0.013}          \\
CLOVER$_{MAMO}$   & \textbf{0.711\std{0.004}} & \textbf{0.786\std{0.002}} & \textbf{0.612\std{0.044}} & \textbf{0.036\std{0.017}} & \textbf{0.039\std{0.011}} \\ \hline
\end{tabular}}
\end{table}

\subsection{Fairness in Recommendation}
Fairness has increasingly become one of the most important topics in machine learning. Recently, there has been a small amount of work focused on fairness in recommendation tasks. The fairness problem in the recommender system is sophisticated because there are multi-stakeholders and various kinds of fairness measures. As such, most of the previous works take a pretty different perspective. In our work, we are concerned about fairness on the user side due to the personalized requirements of the recommendation system \cite{li2021user}. \citet{yao2017beyond} proposed four new metrics for the group fairness in the recommendation and added corresponding constraints to the learning objective to explore the fairness in collaborative filtering recommender systems. \citet{beutel2019fairness} leveraged pairwise comparisons to measure the fairness and designed a pairwise regularization training approach to improve recommender system fairness. \citet{li2021user} studied fairness in recommendation algorithms from the user perspective and applied the fairness constrained re-ranking method to mitigate the group unfairness. \citet{wu2021learning} leveraged the higher-order information in the interaction graph to mitigate the individual fairness problem. \citet{li2021towards} first explored the concept of counterfactual fairness in the recommendation by generating feature-independent user embeddings to satisfy the counterfactual fairness requirements. There are also related works \cite{zhang2021causal, wei2021model} explored the bias of users in the recommender system. \citet{zhu2021fairness} explored the activity fairness problem of new items and designed a learnable post-processing framework to enhance fairness. However, their method is not designed for meta-learning and is not applicable for user-oriented fairness. To the best of our knowledge, our work is the first to explicitly consider different kinds of fairness and the fairness in the meta-learned recommender systems. In our work, we are concerned about the unfair issues on the user side and design a comprehensive framework to mitigate the unfairness issue with adversarial training.  

\subsection{Fairness in Meta Learning}
With the development of meta-learning, a few works have started to explore the fairness problem of it. \citet{slack2020fairness,zhao2020fair} proposed to add group regularization terms in the optimization process of meta-learning. \citet{zhao2020primal} also designed a novel fair meta-learning framework with a task-specific group soft fairness constraint. 
\citet{zhao2021fairness} then extended their previous work to study the problem of online fairness-aware meta-learning to deal with non-stationary task distributions using the long-term fairness constraints. However, these methods are not aimed at recommender systems and didn't consider the properties of recommender systems, thus not applicable for user-oriented fairness. Moreover, these methods are built based on adding group distribution regularized constraints on the inner and outer loop of the meta-learning algorithms. Therefore they can't be applied in the field of recommender systems, since in each adaptation, we can only access the data from every single user. As far as we know, we are the first to explore the fairness issue in the meta-learned recommender system.

%% file: 5-conclusion.tex
\section{Conclusion}
\label{section:conclusion}
In this paper, we present the first fairness view for the meta-learned recommender systems with cold-start. We systematically illustrate the fairness issues in the recommender systems. Then we propose the concept of comprehensive fairness, and formulate it as the adversarial learning problem. A carefully designed framework, \model, is proposed to enable fair adversarial learning in the meta-learned recommender system. Extensive experiments on three real-world data sets demonstrate the effectiveness of \model, which outperforms a set of strong baselines. In the future, we would like to design fairness metrics for multi-class sensitive attributes and explore the fairness issues within the combination of multiple sensitive attributes. We are also interested in considering fairness from a user-item interaction graph perspective.

%% file: 6-appendix.tex
\appendix
\section{Datasets}
\label{appendix:datasets}
To verify the effectiveness of our proposed method, we conducted experiments on three publicly accessible recommender system datasets. We follow \cite{lee2019melu, beigi2020privacy} and use ML-1M , BookCrossing and  ML-100K as our experimental dataset. ML-100K. ML-1M and ML-100K datasets are both from the Movielens Collection\footnote{The MovieLens dataset: \url{https://grouplens.org/datasets/movielens/}}, which consists of the movie rating data from the users who joined the MovieLens website. BookCrossing\footnote{The BookCrossing dataset: \url{http://www2.informatik.uni-freiburg.de/~cziegler/BX/}} is another widely used benchmark for recommender system. It includes the book rating data released on the web. These three datasets all provide the basic information of users and items, as well as the sensitive information of users which can be used for fairness evaluation. We use all the available information for performing recommendations. For attributes with single value, we encode them into one-hot vector for following usage. For attributes with multiple values, we first encode each value into one-hot vector and get embedding of them. Then we average the embedding to obtain the embedding of this attribute.

\section{Metrics}
\label{appendix:metrics}
For evaluating both the recommendation and fairness performance, we adopt the following metrics:
\begin{itemize}[leftmargin=*]
\item \textbf{MAE} (Mean Absolute Error): To measure the errors between prediction and ground-truth:
$$MAE=\frac{1}{|U^f|} \sum_{u \in U^f} \frac{1}{|D^q_u|} \sum_{i \in D^e_u}\left|y_{ui}-\hat{y}_{ui}\right|$$
\item \textbf{NDCG}: The normalized DCG to measure the average performance of a search engine's top-K ranking algorithm.
$$
\begin{aligned}
n D C G_{k} &=\frac{1}{|U^f|} \sum_{u \in U^f} \frac{D C G_{k}^{u}}{I D C G_{k}^{u}} \\
D C G_{k}^{u} &=\sum_{r=1}^{k} \frac{2^{y_{u}^r}-1}{\log _{2}(1+r)}
\end{aligned}
$$
where $y_{u}^r$ is the real rating of user $u$ for the $r$-th ranked item. $DCG_k$ calculates the top-k actual rating values sorted by predicted rating values; and $IDCG_k$ calculates the top-k sorted actual rating values, which is the best possible value. Following \cite{dong2020mamo}, we set $k$ to 3 to measure the performance.

\end{itemize} 
The above metrics measure the performance of the recommender, next, we will introduce the metrics for quantifying the performance of comprehensive fairness in the recommender system.
\begin{itemize}[leftmargin=*]
\item \textbf{AUC} (Area Under the Curve): The curve is created by plotting the true positive rate against the false-positive rate at various threshold settings.
\item \textbf{CF} (Counterfactual Fairness): To measure the difference of predictions by perturbing the sensitive attribute. We denote the sensitive attribute of user $u$ as $a_u$ (e.g., female), any other attributes as $x_u$. $a_u'$ is the other value attainable of the sensitive attribute (e.g., male).
\begin{equation*}
\setlength{\abovedisplayskip}{5pt}
\setlength{\belowdisplayskip}{5pt}
  CF=\frac{1}{|U^f|} \sum_{u \in U^f} \frac{1}{|D^q_u|} \sum_{i \in D^q_u}\left|\hat{y}_{ui}(x_u,a_u)-\hat{y}_{ui}(x_u,a_u')\right|
\end{equation*}
\item \textbf{GF} (Group Fairness): Same as the definition in \ref{subsection:fairness}, we consider GF as the difference of recommender performance (MAE) between users of diverse sensitive attribute.
$$
    GF=|\frac{1}{|A_1|}\sum_{u_1\in A_1} R(u_1) - \frac{1}{|A_2|}\sum_{u_2\in A_2} R(u_2)|
$$
where $A_1$ and $A_2$ are two user groups splited by the sensitive attribute $A$.
\end{itemize}

\section{Baselines}
\label{appendix: baselines}
We compare our method with the following baselines:
\begin{itemize}[leftmargin=*]
\item \textbf{PPR \cite{park2009pairwise}} (Pairwise Preference Regression): It estimates user preferences via bilinear regression.
user and item content vectors.
\item \textbf{Wide \& Deep \cite{cheng2016wide}}: It predicts whether the
user likes an item via a deep neural network. Here the neural network architecture is the same as MELU.
\item \textbf{DropoutNet \cite{volkovs2017dropoutnet}}: It combines the dropout technique with a deep neural network to learn effective features of the input to solve the cold-start problem.
\item \textbf{NLBA \cite{vartak2017meta}}: It learns a cold-start neural network recommender, where all weights (output and hidden) in it are constant across users, while the biases of output and hidden units are adapted per user. Note that though the algorithm is called meta, it only adapts the bias for each user and does not include the meta update.
\item \textbf{MELU \cite{lee2019melu}}: A personalized user preference estimation model based on meta-learning that can rapidly adapt to new users. We adopt this method as our basic model for our framework.
\end{itemize}
The first two baselines are about content-based filtering. The next two baselines are about the traditional cold-start recommender system. For fairness consideration, due to the lack of relevant baselines, we test the following fairness approaches instead based on MELU:
\begin{itemize}[leftmargin=*]
    \item \textbf{BS \cite{koren2009matrix}: } BS learns a biased score from the training stage and then removes the bias in the prediction in the testing stage.
    \item \textbf{Reg \cite{yao2017beyond}: } Reg is a regularization-based approach that directly adds the fairness objective on the loss function. However, this approach can't be directly applied to our problem. Here we try to optimize the fairness objective in the outer loop meta update process. We add both the group fairness and counterfactual fairness objectives in the outer loop loss function.
    \item \textbf{IPW \cite{liang2016causal} }: IPW adopt the reweight-based method \cite{wang2022training} and leverage the standard inverse propensity weight to reweight samples to alleviate the fairness issues.
    \item \textbf{MACR}~\cite{wei2021model}: It provides a general debias framework for recommender system. In our context, it captures the bias related to the sensitive information at the training step and removes the bias at the inference stage.
\end{itemize}
Implementation details and detailed parameter settings of the models can be found in the next section.

\section{Implemented Details}
\label{appendix:implemented details}
We use Pytorch to build our model. We set the dimensions of embedding vectors of all the above models to 64 by default, and use the same recommender loss function for all models for a fair comparison. For Reg method, we test the fairness objective coefficient from 0 to 1 with a step size of 0.1, and is set to 0.5 because it works best. For MACR model, we search the bias elimination parameter $c$ from 20 to 40 with the step size of 2, which is suggested by the authors. The trainable parameters are initialized with the Xavier method. Adam algorithm is used to optimize the model. We set the number of training epochs to 50 and trains 32 user tasks in a batch. During the sampling within each batch, we uniformly sample from all users. The inner learning rate $\alpha$ is set to 1e-2 and the outer loop learning rate $\beta$ is set to 1e-3. For all meta-learned models, we set the number of finetuning steps in the inner loop to 5. The loss function trade-off hyper-parameter $\lambda, \gamma$ are both searched in the range of $\{1e-2,1e-1,1,5\}$ $\lambda$ and $\gamma$ are set to 1 and 0.1 by default.

\begin{table}[t]
\setlength{\abovecaptionskip}{0.cm}
\caption{Effect of hyper-parameter $\lambda$.}
\label{table:lambda}
\scalebox{0.9}{
\begin{tabular}{c|ccccc}
\hline
     & MAE $\downarrow$          & NDCG $\uparrow$           & AUC $\downarrow$          & CF $\downarrow$           & GF $\downarrow$           \\ \hline
1e-2 & 0.735\std{0.002}          & 0.755\std{0.003}          & 0.872\std{0.056}          & 0.092\std{0.019}          & 0.050\std{0.011}          \\
1e-1 & 0.733\std{0.004}          & 0.754\std{0.004}          & 0.679\std{0.052}          & 0.067\std{0.009}          & 0.045\std{0.007}          \\
1    & \textbf{0.731\std{0.005}} & \textbf{0.756\std{0.005}} & \textbf{0.632\std{0.057}} & \textbf{0.044\std{0.014}} & \textbf{0.040\std{0.006}} \\
5    & 0.737\std{0.007}          & 0.754\std{0.004}          & 0.722\std{0.043}          & 0.079\std{0.021}          & 0.048\std{0.006}          \\ \hline
\end{tabular}}
\end{table}

\begin{table}[t]
\setlength{\abovecaptionskip}{0.cm}
\caption{Effect of hyper-parameter $\gamma$.}
\label{table:gamma}
\scalebox{0.9}{
\begin{tabular}{c|ccccc}
\hline
     & MAE $\downarrow$          & NDCG $\uparrow$           & AUC $\downarrow$          & CF $\downarrow$           & GF $\downarrow$           \\ \hline
1e-2 & 0.733\std{0.003}          & 0.754\std{0.004}          & 0.647\std{0.046}          & 0.055\std{0.013}          & 0.044\std{0.011}          \\
1e-1 & \textbf{0.731\std{0.005}} & \textbf{0.756\std{0.005}} & \textbf{0.632\std{0.057}} & \textbf{0.044\std{0.014}} & \textbf{0.040\std{0.006}} \\
1    & 0.734\std{0.005}          & 0.754\std{0.003}          & 0.651\std{0.055}          & 0.060\std{0.020}          & 0.047\std{0.006}          \\
5    & 0.741\std{0.003}          & 0.754\std{0.004}          & 0.941\std{0.017}          & 0.176\std{0.023}          & 0.061\std{0.008}          \\ \hline
\end{tabular}}
\end{table}

\section{Effect of Hyper-parameters (RQ3)}
\label{appendix: hyper-parameters}
As formulated in the loss function, $\lambda$ is the trade-off hyper-parameter that balances the contribution of the recommendation model loss and the individual and counterfactual fairness loss while $\gamma$ is to balance the recommender model loss and the group fairness loss. To investigate the benefits of them, we conduct experiments of \model~  with varying $\lambda$ and $\gamma$ respectively. In particular, we search their values in the range of \{0.01, 0.1, 1, 5\}. When varying one parameter, the other is set as constant. From Table~\ref{table:lambda} and Table~\ref{table:gamma} we have the following findings. We can find our method can improve the recommender and fairness performance in a wide range. As $\lambda$ increases from 1e-2 to 1, the performance of \model~ will become better. A similar trend can be observed by varying $\gamma$ from 1e-2 to 1e-1. However, when $\lambda$ or $\gamma$ surpasses a threshold (1 for $\lambda$), the performance becomes worse. As the hyper-parameters become larger, the training of the recommendation model will be less important and the bi-level meta-learned model will be difficult to converge, which brings the worse results.